# Statistical analysis of motion contrast in optical coherence tomography angiography

Yuxuan Cheng
Li Guo
Cong Pan
Tongtong Lu
Tianyu Hong
Zhihua Ding
Peng Li

# Statistical analysis of motion contrast in optical coherence tomography angiography

**Yuxuan Cheng,[a] Li Guo,[a] Cong Pan,[a] Tongtong Lu,[a] Tianyu Hong,[b] Zhihua Ding,[a] and Peng Li[a,*]**
[a]Zhejiang University, College of Optical Science and Engineering, State Key Laboratory of Modern Optical Instrumentation and the Collaborative Innovation Center for Brain Science, Hangzhou, Zhejiang 310027, China
[b]Zhejiang University, College of Biomedical Engineering and Instrument Science, Hangzhou, Zhejiang 310027, China

**Abstract.** Optical coherence tomography angiography (Angio-OCT), mainly based on the temporal dynamics of OCT scattering signals, has found a range of potential applications in clinical and scientific research. Based on the model of random phasor sums, temporal statistics of the complex-valued OCT signals are mathematically described. Statistical distributions of the amplitude differential and complex differential Angio-OCT signals are derived. The theories are validated through the flow phantom and live animal experiments. Using the model developed, the origin of the motion contrast in Angio-OCT is mathematically explained, and the implications in the improvement of motion contrast are further discussed, including threshold determination and its residual classification error, averaging method, and scanning protocol. The proposed mathematical model of Angio-OCT signals can aid in the optimal design of the system and associated algorithms.  

## 1 Introduction

Optical coherence tomography angiography (Angio-OCT) is capable of contrasting the dynamic blood flow against the static tissue bed with high spatial resolution and high motion sensitivity (down to capillary level) in a depth-resolved manner. Up until now, several Angio-OCT algorithms have been developed for generating the motion contrast.[1–17] Generally, each spatial position should be sampled/imaged several times with a certain time interval by using repeated[10,13,16] or dense[14,18] scanning protocols in the OCT system. Then the temporal changes in amplitude (or intensity),[2,4,10,11] phase,[5,9,19] or complex-value[7,13,17,20] of OCT signals over such a time interval are analyzed with different processing algorithms, such as speckle variance,[2–4] Doppler variance,[7,8,14] phase variance,[5,6] differential calculation,[18] and correlation mapping.[11,21] By circumventing the exogenous contrast injection, such motion-contrast Angio-OCT provides great advantages over conventional flourescence-based angiography.

Knowledge of the statistical properties of Angio-OCT signals would be helpful for further understanding the origin of the motion-contrast and guiding the optimization of the system and associated algorithms. It is well known that the motion-contrast Angio-OCT is mainly based on the temporal dynamics of OCT scattering signals, and the algorithms of differential calculation are widely used for dynamics analysis, including the amplitude differential (AD) and complex differential (CD) algorithms. The temporal statistics of the OCT amplitude signals have been well documented in the literature.[22,23] In a similar way, the temporal statistics of the complex-valued OCT signals can be mathematically described based on the knowledge of statistical optics.[24] In this study, the statistical properties of AD- and CD-Angio-OCT signals are derived in theory, and the implications of the developed statistical model are briefly illustrated.

This work is organized as follows. (1) The temporal statistics of the complex-valued OCT signals were first described and the mathematical statistics of AD- and CD-Angio-OCT were further deduced in Sec. 3. (2) The statistical properties derived in theory were validated through the flow phantom and live animal experiments in Sec. 4. (3) The potential implications of the statistical properties were discussed in Sec. 5.

## 2 Materials and Methods

### 2.1 *Flow Phantom and Animal Preparation*

The flow phantom was made of an agarose gel mixed with ∼5% milk to mimic the static scattering tissue background. A capillary tube with an inner diameter of 0.5 mm was embedded in this tissue-like phantom. A 3% milk solution was pumped into the tube at a constant rate with a syringe pump (KDS 100 series, Stoelting Co., Wood Dale, Illinois) to simulate the flowing blood.

C57BL/6 mice of 8 to 10 weeks old were used in animal experiments. A mouse was anesthetized by intraperitoneal injection of 10% chloral hydrate (4 ml/kg). Its head was fixed in a stereotaxic frame (Stoelting), and the scalp was retracted. The skull was thinned by using a saline-cooled dental drill to generate a window of 3 mm × 3 mm area and to facilitate the optical penetration within the cortex at an 850 nm wavelength. All animals were provided by the Experimental Animal Center and treated with the guidelines of the Institutional Animal Care and Use Committee of Zhejiang University.

*Address all correspondence to: Peng Li, E-mail: peng_li@zju.edu.cn

### 2.2 *System Setup and Scanning Protocol*

The imaging system in this study was built based on a typical configuration of spectral domain OCT (SDOCT). Briefly, the light source is a broadband superluminescent diode with a central wavelength of 850 nm and a full width at half maximum bandwidth of 100 nm, theoretically offering a high axial resolution of $\sim 3.2\ \mu$m in air. The measured lateral resolution is $\sim 15\ \mu$m. A high-speed spectrometer equipped with a fast line scan CMOS camera was used as the detection unit in our system, providing a 120-kHz line scan rate.

In this study, MB-mode scanning protocol was used for dynamic analysis. Each B-scan was formed by 512 A-lines, determining a rate of 190 fps. Therefore, 1000 repeated B-scans were sequentially acquired in the same cross-section within ∼5.3 s, generating a three-dimensional (3-D) OCT data cube $(z, x, n)$, where $n$ is the B-frame index, equivalent to the time dimension. $z$ and $x$ represent the depth and transverse position, respectively, as shown in Fig. 1(a).

### 2.3 *Processing Algorithm*

The depth-resolved complex reflectivity of a scattering sample is reconstructed by performing Fourier transform of the spectral interference fringe signals in SDOCT. The complex-valued OCT signals of the $n$'th repeated B-frame is denoted as $\tilde{A}(z, x, n)$. A map of the amplitude signals $A(z, x, n)$ is used to generate the structural image. The differences of the amplitude $A(z, x, n)$[18] and complex-valued $\tilde{A}(z, x, n)$[17] OCT signals between adjacent B-frames are computed for AD- and CD-Angio-OCT, respectively, as follows:

$$\text{AngioOCT}_{\text{AD}} = a_{\text{AD}}(z, x, n) = A(z, x, n+1) - A(z, x, n), \tag{1}$$

$$\text{AngioOCT}_{\text{CD}} = a_{\text{CD}}(z, x, n) = |\tilde{A}(z, x, n+1) - \tilde{A}(z, x, n)|, \tag{2}$$

where $a_{\text{AD}}$ and $a_{\text{CD}}$ represent the amplitude of AD-Angio-OCT and CD-Angio-OCT signals, respectively. Typically, the absolute value of $a_{\text{AD}}$ is used in the final angiograms in AD-Angio-OCT. The adjacent B-frames are acquired at the same cross section with a certain time interval ($t$). Then thresholds are used to identify the dynamic areas. Due to the bulk motion, prior to the subtraction operation in Eq. (2), the global phase fluctuations are determined and compensated by a histogram-based phase selecting process.[9,16,25–27]

In this study, the measured data were statistically analyzed with the histogram, and then compared with the statistical model proposed in theory. R-square ($R^2$) statistic was measured to evaluate how well the experimental outcome fit the theoretical prediction.[28]

## 3 Theory

The flow chart of the mathematical analysis is depicted in Fig. 1. Because the complex-valued OCT signal is used in the CD-Angio-OCT, the temporal distribution of the complex-valued OCT signals was briefly described [Fig. 1(b)]. Then the mathematical statistics of Angio-OCT signals were deduced in theory [Fig. 1(c)]. The following derivations are mainly based on the knowledge of random phasor sums and transformations of random variables which have been well elaborated in Ref. 24. We focus on the physical plausibility and the associated assumptions used for solving OCT problems.

### 3.1 *Temporal Statistics in Optical Coherence Tomography*

For brevity, at a given position, the complex-valued OCT signals $\tilde{A}(z, x)$ are denoted by a phasor $a_{d/s} \exp(j\theta_{d/s})$

$$\tilde{A}(z, x) = a_{d/s} \exp(j\theta_{d/s}), \tag{3}$$

where the subscripts $d$ and $s$ represent the dynamic and static signals, respectively. As a result of the coherence gating used in OCT, the phasor $a_{d/s} \exp(j\theta_{d/s})$ is a complex addition of many small phasors, arising from a collection of small scatters that are distributed within the OCT voxel of interest.[29]

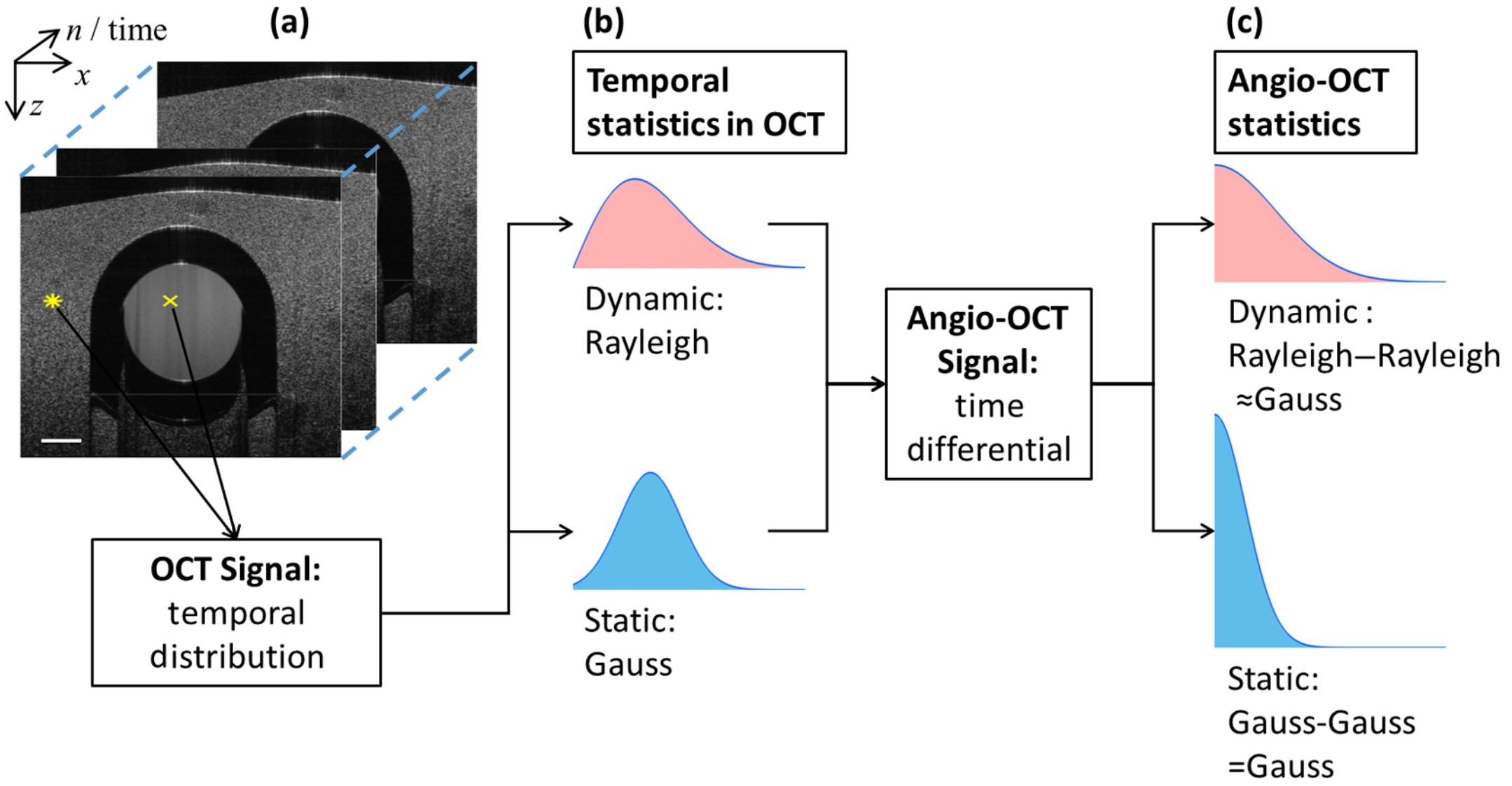

**Fig. 1** Flow chart of the statistical analysis in optical coherence tomography angiography (Angio-OCT). Amplitude differential (AD) method is used as an example for illustration. (a) Three-dimensional (3-D) OCT data cube $(z, x, n)$. (b) Temporal distribution of the complex-valued OCT signals. (c) Statistical distribution of the AD-Angio-OCT signals.

In the dynamic regions, the small scatters are mainly contributed by the moving red blood cells (RBCs). The corresponding light scattering signals are time variant. Thus, the resultant phasor $a_d \exp(j\theta_d)$ can be regarded as a complex sum of a large number of small, random phasors, usually referred to as random phasor sum.

$$a_d \exp(j\theta_d) = \sum_{l=1}^{M} \beta_l \exp(j\varphi_l) \quad = r_d + j \cdot i_d, \tag{4}$$

where $\beta$, $\varphi$, and $M$ are the amplitude, phase, and total number of the small independent phasor in the dynamic regions, respectively. $r_d$ and $i_d$ are the real and imaginary parts of the resultant phasor, respectively. The properties of random phasor sum have been detailed in Ref. 24. Briefly, in the limit of a very large $M$, the resultant phasor $a_d \exp(j\theta_d)$ is a circular complex Gaussian random variable.[24] And the joint probability density function (PDF) of $r_d$ and $i_d$ is

$$f_{R_d I_d}(r_d, i_d) = \frac{1}{2\pi\sigma_d^2} \exp\left(-\frac{r_d^2 + i_d^2}{2\sigma_d^2}\right), \tag{5}$$

where $\sigma_d^2$ represents the variance of the real and imaginary parts. Using the transformation $r_d = a_d \cos\theta_d$, $i_d = a_d \sin\theta_d$, the temporal PDF of amplitude $a_d$ is derived.[24]

$$f_{A_d}(a_d) = \begin{cases} \frac{a_d}{\sigma_d^2} \exp\left(-\frac{a_d^2}{2\sigma_d^2}\right) & a_d \ge 0 \\ 0 & \text{otherwise.} \end{cases} \tag{6}$$

Therefore, in the dynamic regions, the amplitude of the OCT signals obeys a Rayleigh distribution with mean $\sqrt{\pi/2}\sigma_d$ and variance $(2 - \pi/2)\sigma_d^2$, which is in good agreement with the literature.[30]

In the static regions, the light scattering signals are time invariant, and can be regarded as a strong, constant phasor. In this case, the system noise becomes the primary random contribution, and the phasor $a_{\text{noise}} \exp(j\theta_{\text{noise}})$ of the noise can be regarded as a weak random phasor sum

$$a_{\text{noise}} \exp(j\theta_{\text{noise}}) = \sum_{l=1}^{N} \alpha_l \exp(j\phi_l) = r_{\text{noise}} + j \cdot i_{\text{noise}}, \tag{7}$$

where $\alpha$, $\varphi$, and $N$ are the amplitude, phase, and total number of the small independent phasor in the static regions, respectively. $r_{\text{noise}}$ and $i_{\text{noise}}$ are the real and imaginary parts of the random phasor sum, respectively. The weak random phasor sum is a circular complex Gaussian random variable with zero mean and standard deviation $\sigma_s$. In most situations of interest, the OCT signal $C$ is much stronger than the system noise. Thus, the resultant phasor $a_s \exp(j\theta_s)$ equals a strong constant phasor $C$ plus a weak random phasor sum, as follows:

$$\begin{aligned} a_s \exp(j\theta_s) &= C + a_{\text{noise}} \exp(j\theta_{\text{noise}}) \\ &= (C + r_{\text{noise}}) + j \cdot i_{\text{noise}} = r_s + j \cdot i_s, \end{aligned} \tag{8}$$

where $r_s$ and $i_s$ are the real and imaginary parts of the resultant phasor, respectively. The joint PDF of $r_s$ and $i_s$ follows a two-dimensional Gaussian distribution[24]

$$f_{R_s I_s}(r_s, i_s) = \frac{1}{2\pi\sigma_s^2} \exp\left[-\frac{(r_s - C)^2 + i_s^2}{2\sigma_s^2}\right], \tag{9}$$

with $C \gg \sigma_s$, the approximation is made that

$$a_s \approx C + r_{\text{noise}}. \tag{10}$$

Variations in the amplitude $a_s$ are primarily caused by the real part $r_{\text{noise}}$ of the weak random phasor sum, thus we have

$$f_{A_s}(a_s) \cong \begin{cases} \frac{1}{\sqrt{2\pi}\sigma_s} \exp\left[-\frac{(a_s - C)^2}{2\sigma_s^2}\right] & a_s \ge 0 \\ 0 & \text{otherwise.} \end{cases} \tag{11}$$

In the static regions, the amplitude of the OCT signals obeys a Gaussian distribution with mean $C$ and variance $\sigma_s^2$, which is also in agreement with the literature.[30]

## 3.2 *Angio-Optical Coherence Tomography Statistics*

### 3.2.1 *Amplitude differential-angio-optical coherence tomography*

In AD-Angio-OCT, substituting Eq. (3) into Eq. (1) yields

$$\text{AngioOCT}_{\text{AD}} = a_{\text{AD}d/s}(n) = a_{d/s}(n+1) - a_{d/s}(n). \tag{12}$$

In the dynamic regions, we have

$$a_{\text{AD}d}(n) = a_d(n+1) - a_d(n). \tag{13}$$

Due to the moving of RBC, the variables $a_d(n+1)$ and $a_d(n)$ which represent the amplitude of the scattering light can be regarded to be independent and random, and they follow the same but independent Rayleigh distribution, as described by Eq. (6). The statistics of the random variable $a_{\text{AD}d}$ can be derived as follows:[24]

$$\begin{aligned} f_{A_{\text{AD}d}}(a_{\text{AD}d}) &= -\int_{-\infty}^{+\infty} f_{A_d}(w - a_{\text{AD}d}) f_{A_d}(w) \mathrm{d}w \\ &= \frac{\sqrt{\pi}}{4\sigma_d}\left(1 - \frac{a_{\text{AD}d}^2}{2\sigma_d^2}\right)\left[1 - \text{erf}\left(\frac{a_{\text{AD}d}}{2\sigma_d}\right)\right] \\ &\quad \times \exp\left(-\frac{a_{\text{AD}d}^2}{4\sigma_d^2}\right) + \frac{a_{\text{AD}d}}{4\sigma_d^2} \exp\left(-\frac{a_{\text{AD}d}^2}{2\sigma_d^2}\right), \end{aligned} \tag{14}$$

where the $w$ is an intermediate variable. The erf is the Gauss error function. Unfortunately, we do not have a simplified analytic expression of Eq. (14). According to numerical simulation in MATLAB® as shown in Fig. 2, Eq. (14) is extremely close to a Gaussian distribution with zero mean and variance $\sigma_d^2$:

$$f_{A_{\text{AD}d}}(a_{\text{AD}d}) \approx \frac{1}{\sqrt{2\pi}\sigma_d} \exp\left(-\frac{a_{\text{AD}d}^2}{2\sigma_d^2}\right). \tag{15}$$

And the absolute value $|a_{\text{AD}d}|$ follows a truncated Gaussian distribution with mean $\sigma_d/\sqrt{2\pi}$ and variance $(1 - 2/\pi)\sigma_d^2$:

$$f_{|A_{\text{AD}d}|}(|a_{\text{AD}d}|) \approx \begin{cases} \frac{2}{\sqrt{2\pi}\sigma_d} \exp\left(-\frac{|a_{\text{AD}d}|^2}{2\sigma_d^2}\right) & a_{\text{AD}d} \ge 0 \\ 0 & a_{\text{AD}d} < 0. \end{cases} \tag{16}$$

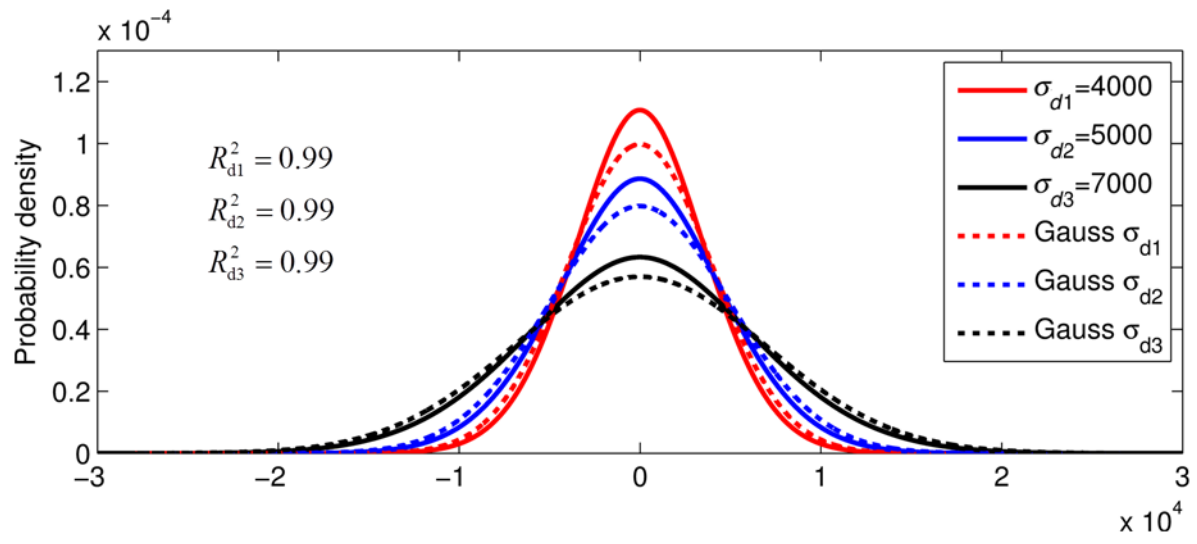

**Fig. 2** Numerical simulation of AD-Angio-OCT signals in the dynamic region. The solid lines are the numerical simulation of Eq. (14) with different variances. The dash lines are the corresponding Gaussian distributions of Eq. (15). The R-square value ($R^2$) is 0.99.

In the static regions, the OCT scattering signal $C$ is time invariant, and remains constant in all the B-frames for a given spatial point. Substituting Eq. (10) into Eq. (12), we have

$$a_{\mathrm{ADs}}(n) = a_s(n+1) - a_s(n) \approx r_{\mathrm{noise}}(n+1) - r_{\mathrm{noise}}(n). \quad (17)$$

The $r_{\mathrm{noise}}(n+1)$ and $r_{\mathrm{noise}}(n)$ are random variables, which obey the same but independent Gaussian distribution [Eq. (9)]:

$$f_{R_{\mathrm{noise}}}(r_{\mathrm{noise}}) = \frac{1}{\sqrt{2\pi}\sigma_s} \exp\left(-\frac{r_{\mathrm{noise}}^2}{2\sigma_s^2}\right). \quad (18)$$

The statistics of the random variable $a_{\mathrm{ADs}}$ can be deduced by a transformation of random variables:

$$f_{A_{\mathrm{ADs}}}(a_{\mathrm{ADs}}) = -\int_{-\infty}^{+\infty} f_{R_{\mathrm{noise}}}(v - a_{\mathrm{ADs}}) f_{R_{\mathrm{noise}}}(v)\mathrm{d}v = \frac{1}{\sqrt{2\pi}\left(\sqrt{2}\sigma_s\right)} \exp\left[-\frac{a_{\mathrm{ADs}}^2}{2\left(\sqrt{2}\sigma_s\right)^2}\right], \quad (19)$$

where $v$ is an intermediate variable. It is a Gaussian distribution with zero mean and variance $2\sigma_s^2$. Similar to Eq. (16), the absolute value $|a_{\mathrm{ADs}}|$ obeys a truncated Gaussian distribution with mean $2\sigma_s/\sqrt{\pi}$ and variance $(2 - 4/\pi)\sigma_s^2$:

$$f_{|A_{\mathrm{ADs}}|}(|a_{\mathrm{ADs}}|) = \begin{cases} \frac{2}{\sqrt{2\pi}\left(\sqrt{2}\sigma_s\right)} \exp\left[-\frac{|a_{\mathrm{ADs}}|^2}{2\left(\sqrt{2}\sigma_s\right)^2}\right] & a_{\mathrm{ADs}} \geq 0 \\ 0 & a_{\mathrm{ADs}} < 0. \end{cases} \quad (20)$$

### 3.2.2 *Complex differential-angio-optical coherence tomography*

In the CD-Angio-OCT, Eq. (2) can be rewritten as

$$\mathrm{AngioOCT_{CD}} = a_{\mathrm{CD}d/s}(n) = |a_{d/s}(n+1)\exp[j\theta_{d/s}(n+1)] - a_{d/s}(n)\exp[j\theta_{d/s}(n)]| = \sqrt{[r_{d/s}(n+1) - r_{d/s}(n)]^2 + [i_{d/s}(n+1) - i_{d/s}(n)]^2}. \quad (21)$$

Here, we define four intermediate random variables $\xi_{d/s}$ and $\eta_{d/s}$:

$$\xi_{d/s} = r_{d/s}(n+1) - r_{d/s}(n) = \begin{cases} r_d(n+1) - r_d(n) & \text{dynamic region} \\ r_{\mathrm{noise}}(n+1) - r_{\mathrm{noise}}(n) & \text{static region,} \end{cases} \quad (22)$$

and

$$\eta_{d/s} = i_{d/s}(n+1) - i_{d/s}(n) = \begin{cases} i_d(n+1) - i_d(n) & \text{dynamic region} \\ i_{\mathrm{noise}}(n+1) - i_{\mathrm{noise}}(n) & \text{static region.} \end{cases} \quad (23)$$

Referring to Eq. (5), $r_d(n+1)$, $r_d(n)$, $i_d(n+1)$, and $i_d(n)$ follow the same but independent Gaussian distribution. According to Eqs. (17)–(19), the subtraction of two independent Gaussian distributions remains Gaussian with a modified variance. Thus, the random variables $\xi_d$ and $\eta_d$ have the same but independent Gaussian distribution with zero mean and variance $2\sigma_d^2$. Referring to the derivation from Eqs. (5) to (6), in the dynamic regions, the amplitude $a_{\mathrm{CD}d}$ obeys a Rayleigh distribution with mean $\sqrt{\pi}\sigma_d$ and variance $(4-\pi)\sigma_d^2$, as follows:

$$f_{A_{\mathrm{CD}d}}(a_{\mathrm{CD}d}) = \begin{cases} \frac{a_{\mathrm{CD}d}}{2\sigma_d^2} \exp\left(-\frac{a_{\mathrm{CD}d}^2}{4\sigma_d^2}\right) & a_{\mathrm{CD}d} \geq 0 \\ 0 & a_{\mathrm{CD}d} < 0 \end{cases}. \quad (24)$$

Similarly, in the static regions, the statistics of the variable $a_{\mathrm{CD}s}$ obey a Rayleigh distribution with mean $\sqrt{\pi}\sigma_s$ and variance $(4-\pi)\sigma_s^2$, as follows:

$$f_{A_{\mathrm{CD}s}}(a_{\mathrm{CD}s}) = \begin{cases} \frac{a_{\mathrm{CD}s}}{2\sigma_s^2} \exp\left(-\frac{a_{\mathrm{CD}s}^2}{4\sigma_s^2}\right) & a_{\mathrm{CD}s} \geq 0 \\ 0 & a_{\mathrm{CD}s} < 0 \end{cases}. \quad (25)$$

In this section, the mathematical statistics of the AD-Angio-OCT and CD-Angio-OCT were further deduced. According to Eqs. (16), (20), (24), and (25), the Angio-OCT statistics depend on the variances of the OCT statistics, i.e., $\sigma_d^2$ and $\sigma_s^2$. As a summary, the OCT and Angio-OCT statistics of the dynamic and static signals were tabulated, as shown in Table 1.

## 4 Experimental Validation

### 4.1 *Flow Phantom Imaging*

Figure 3(a) is a representative OCT structural cross section of the flow phantom. The transparent tube is clearly visualized. The regions inside and outside the tube correspond to the dynamic fluid and the static solid gel, respectively. Figures 3(b) and 3(c) show the corresponding cross-sectional angiograms of AD- and CD-Angio-OCT, respectively. As indicated in Fig. 3(a), in the recorded cross section, two spatial points were randomly selected from the regions of static solid-gel $(z_s, x_s)$ and dynamic fluid $(z_d, x_d)$, respectively, and used for the statistical analysis of the OCT amplitude signals and the Angio-OCT signals.

The temporal statistics of the OCT amplitude signals at the selected points are reported in Fig. 4. The histograms present the statistical distributions of the measured data. The variances $\sigma_d^2$ and $\sigma_s^2$ are computed from the measured data. Substituting the variances $\sigma_d^2$ and $\sigma_s^2$ into Eqs. (6) and (11) yields the theoretical predictions of OCT statistics, as plotted by the dashed curves in Fig. 4. The high R-square statistics ($R^2 > 0.95$) indicate the

**Table 1** Statistics in optical coherence tomography and Angio-OCT.

| | OCT Amplitude $a_{d/s}$ | AD-Angio-OCT $\lvert a_{\mathrm{AD}d/s}\rvert$ | CD-Angio-OCT $a_{\mathrm{CD}d/s}$ |
|---|---|---|---|
| Dynamic region | Rayleigh: $\frac{a_d}{\sigma_d^2}\exp\left(-\frac{a_d}{2\sigma_d^2}\right)$ | Truncated Gauss: $\frac{2}{\sqrt{2\pi}\sigma_d}\exp\left(-\frac{\lvert a_{\mathrm{AD}d}\rvert^2}{2\sigma_d^2}\right)$ | Rayleigh: $\frac{a_{\mathrm{CD}d}}{2\sigma_d^2}\exp\left(-\frac{a_{\mathrm{CD}d}^2}{4\sigma_d^2}\right)$ |
| Static region | Gauss: $\frac{1}{\sqrt{2\pi}\sigma_s}\exp\left[-\frac{(a_s-C)^2}{2\sigma_s^2}\right]$ | Gauss: $\frac{2}{\sqrt{2\pi}(\sqrt{2}\sigma_s)}\exp\left[-\frac{\lvert a_{\mathrm{AD}s}\rvert^2}{2(\sqrt{2}\sigma_s)^2}\right]$ | Rayleigh: $\frac{a_{\mathrm{CD}s}}{2\sigma_s^2}\exp\left(-\frac{a_{\mathrm{CD}s}^2}{4\sigma_s^2}\right)$ |

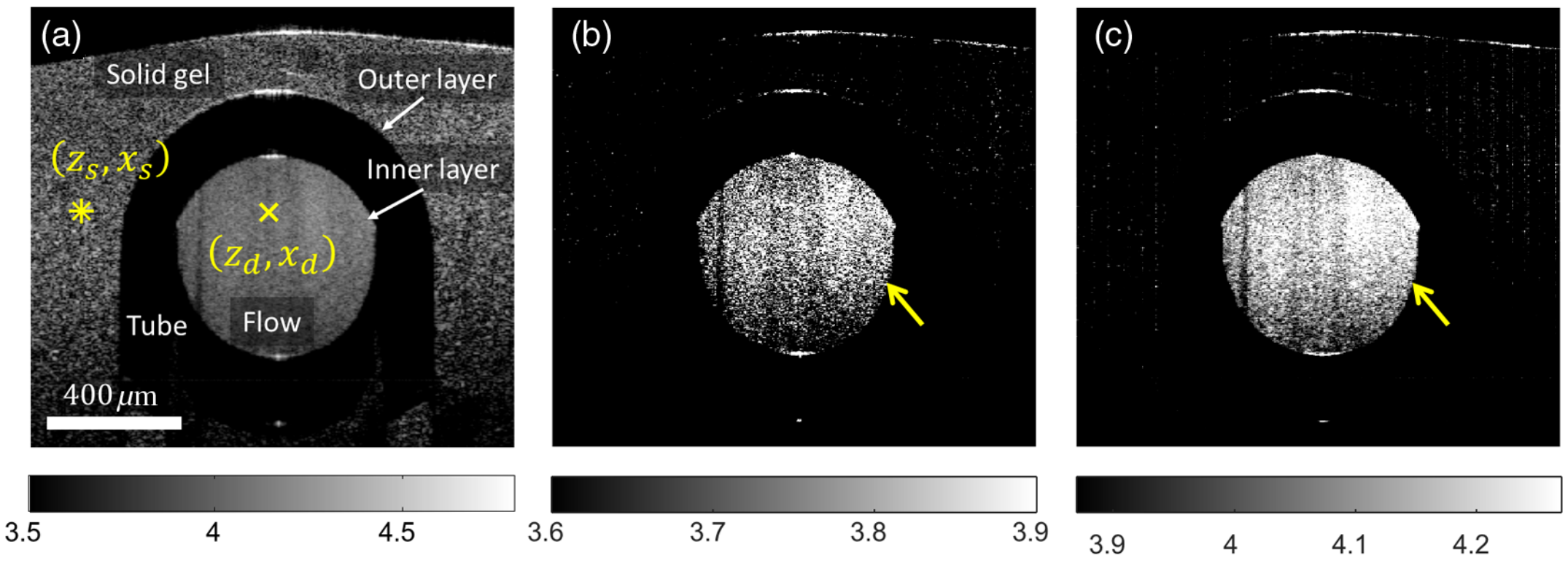

**Fig. 3** (a) Representative structural cross section of flow phantom displayed in log scale. The corresponding cross-sectional angiograms (b) AD-Angio-OCT and (c) CD-Angio-OCT. The asterisk and cross indicate the selected two points from the static and dynamic regions, respectively. (b, c) The yellow arrows indicate the different performance between AD and CD algorithms. Four adjacent frames are averaged for presentation.

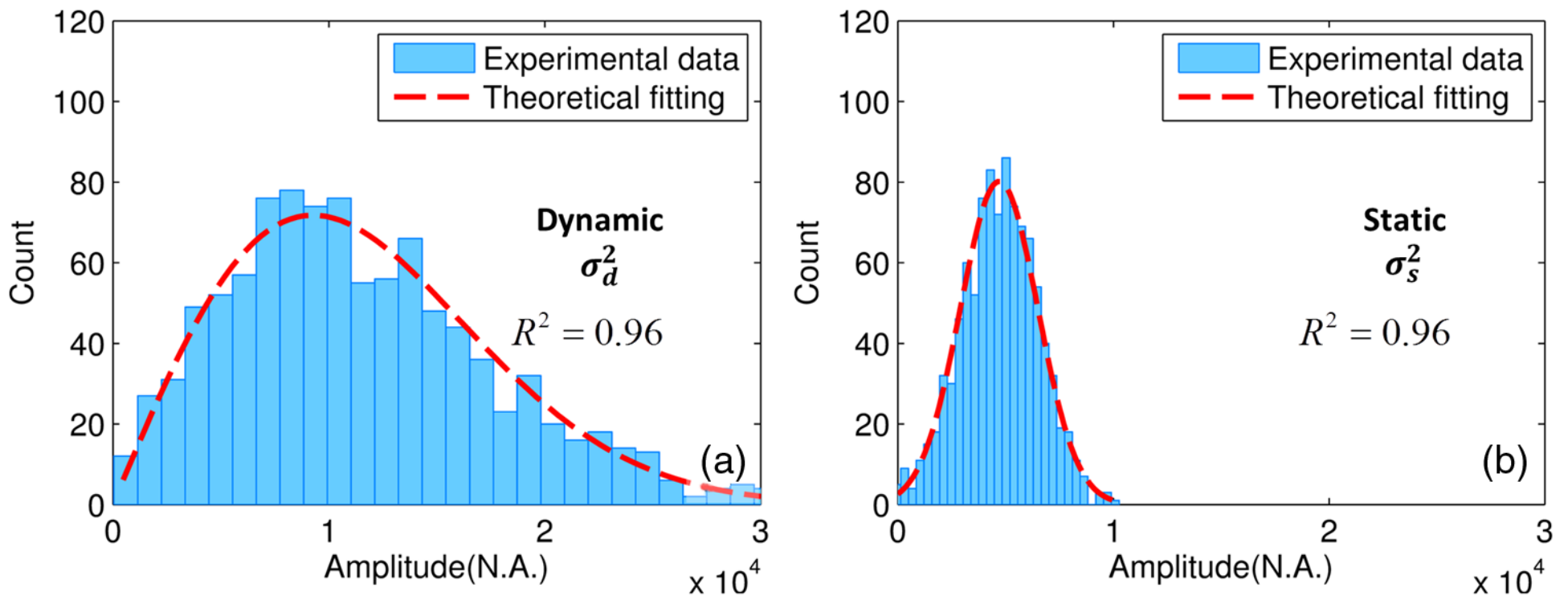

**Fig. 4** Temporal statistics of the OCT amplitude signals at the (a) dynamic and (b) static positions, which are indicated by the cross and asterisk in Fig. 3(a), respectively. The histograms are the statistical distributions of the measured data. The dashed curves correspond to the theoretical predictions. The high R-square statistics ($R^2 > 0.95$) indicate that the experimental outcome matches well with the theoretical model. The variances $\sigma_d^2$ and $\sigma_s^2$ are computed from the measured OCT data.

well-matched agreement between the experimental outcome and the theoretical model.

Figure 5 reports the statistical analysis of the Angio-OCT signals. Figures 5(a) and 5(b) correspond to the dynamic and static regions in AD-Angio-OCT, respectively. Figures 5(c) and 5(d) correspond to the dynamic and static regions in CD-Angio-OCT, respectively. The histograms and solid curves show the experimentally measured and theoretically predicted distributions of the Angio-OCT signals, respectively. The theoretical predictions are generated by substituting the variances $\sigma_d^2$ and $\sigma_s^2$ into Eqs. (16), (20), (24), and (25). The R-square statistics ($R^2$) are higher than 0.95, indicating good agreement between the experimental outcome and the theoretical prediction.

## 4.2 *In Vivo Brain Imaging*

Figure 6(a) is a representative structural cross section of the mouse brain *in vivo*, in which it is challenging to discriminate the dynamic blood flow from the static tissue bed. Figures 6(b) and 6(c) show the corresponding cross-sectional angiograms of AD-Angio-OCT and CD-Angio-OCT, respectively. Similar to

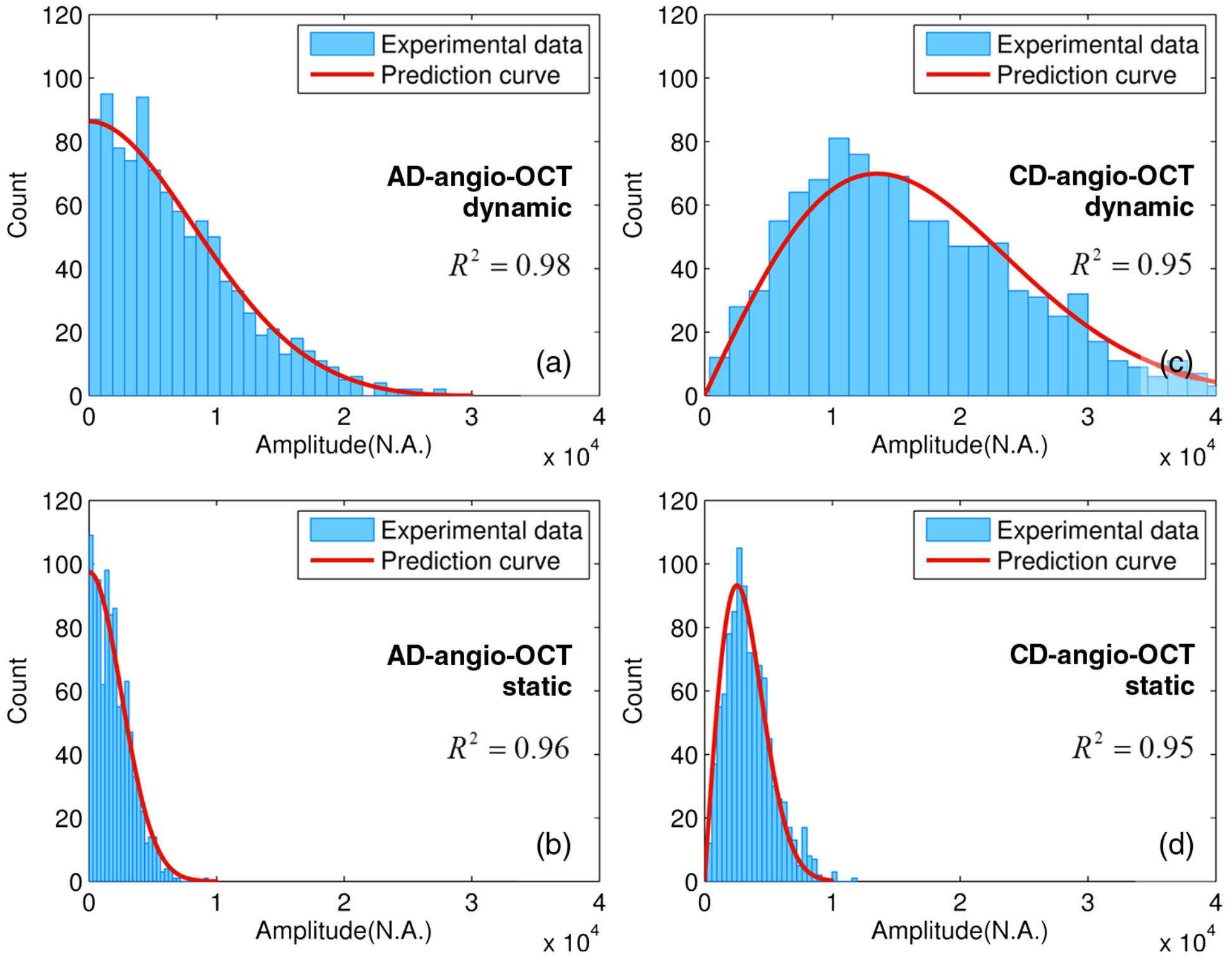

**Fig. 5** Statistics of the Angio-OCT signals in flow phantom imaging. (a, b)The statistics of the dynamic and static signals in AD-Angio-OCT, respectively. (c, d) The statistics of the dynamic and static signals in CD-Angio-OCT, respectively. The R-square statistics ($R^2$) are higher than 0.95.

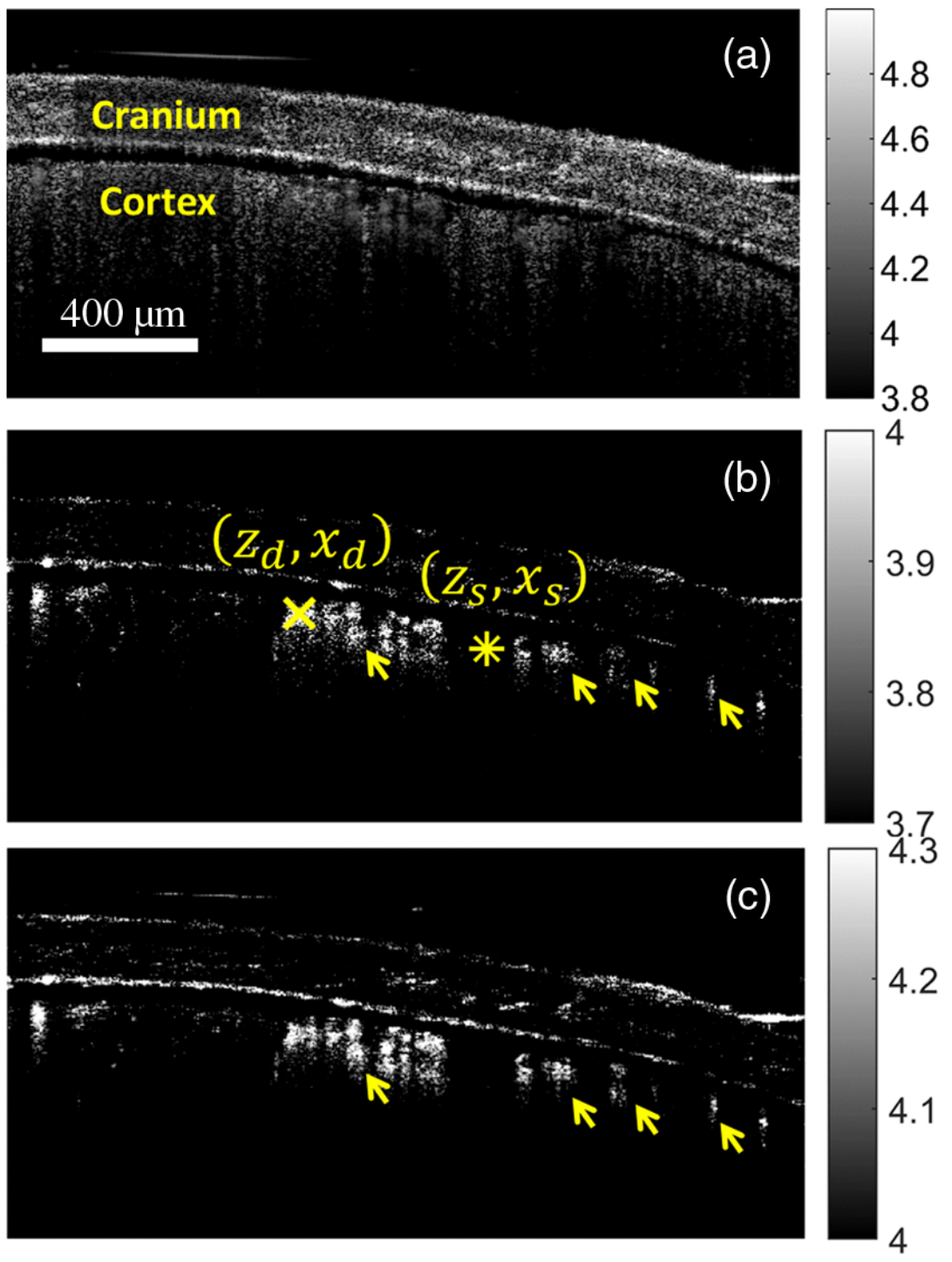

**Fig. 6** (a) Representative structural cross section of mouse brain *in vivo* displayed in log scale. (a, b) The corresponding cross-sectional angiograms are produced by AD-Angio-OCT and CD-Angio-OCT algorithms, respectively. The asterisk and cross indicate the selected two points from the static tissue bed and the dynamic blood flow, respectively. (b, c) The yellow arrows indicate the different performance between AD and CD algorithms. Four adjacent frames are averaged for presentation.

the phantom experiment, in the recorded cross section, two spatial positions were randomly selected from the regions of static tissue bed $(z_s, x_s)$ and dynamic blood flow $(z_d, x_d)$, respectively, for the following statistical analysis. The variances $\sigma_d^2$ and $\sigma_s^2$ are computed from the measured data, and then substituted into Eqs. (6), (11), (16), (20), (24), and (25) for predicting the OCT and Angio-OCT statistics.

The temporal statistics of the OCT amplitude signals at the selected static $(z_s, x_s)$ and dynamic $(z_d, x_d)$ positions are reported in Fig. 7. The histograms show the experimentally measured distributions. The dashed curves in Fig. 7 correspond to the theoretical predictions. The R-square statistics ($R^2$) are higher than 0.95.

Figure 8 reports the temporal statistics of the Angio-OCT signals of the *in vivo* experiment. Figures 8(a) and 8(b) correspond to the blood flow and static tissue bed in AD-Angio-OCT, respectively. Figures 8(c) and 8(d) correspond to the blood flow and static tissue bed in CD-Angio-OCT, respectively. The histograms and the solid curves show the experimentally measured and theoretically predicted distributions of the Angio-OCT signals, respectively. The R-square statistics ($R^2$) are higher than 0.95. The experimental data match well with the theoretical predictions of the proposed statistical models.

## 5 Discussion

Based on the model of random phasor sums, the temporal statistics of the complex-valued OCT signals were mathematically described, as expressed in Eqs. (5), (6), (9), and (11). The dynamic and static signals exhibit intrinsic differences in the statistics of OCT amplitude. Despite the differences, the distributions of the dynamic and static signals still have a large overlap (referring to the error areas in Fig. 9), making it challenging to directly separate the dynamic blood flow from the static tissue bed in the OCT structural image [Figs. 3(a) and 6(a)]. In Angio-OCT, extra processing algorithms, such as the widely used AD and CD algorithms, are applied to the original OCT signals. Using mathematical transformations and reasonable approximations, the statistics of the AD-Angio-OCT and CD-Angio-OCT signals were further derived. In Angio-OCT, the overlap between the distributions of dynamic and static signals is greatly reduced [referring to the error areas in Figs. 9(b) and 9(c)]. The origin of motion–contrast in Angio-OCT has been mathematically explained in this work.

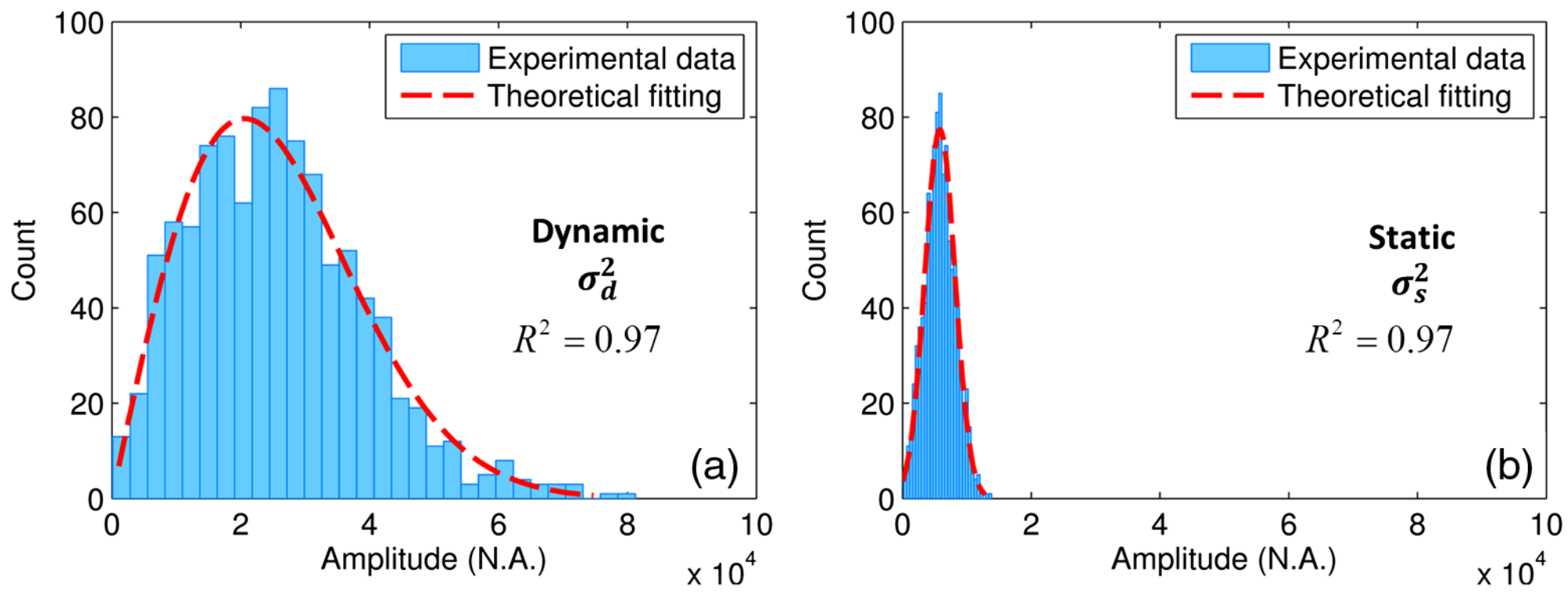

**Fig. 7** Temporal statistics of the OCT amplitude signals from (a) dynamic blood flow and (b) the static tissue bed in mouse brain *in vivo*, which are marked by the cross and asterisk in Fig. 6(a), respectively. The histograms are the distributions of the experimental data. The dashed curves in (a) and (b) correspond to the theoretical fitting using Eqs. (6) and (11). The R-square statistics ($R^2$) are higher than 0.95. The variances $\sigma_d^2$ and $\sigma_s^2$ are computed from the measured OCT data.

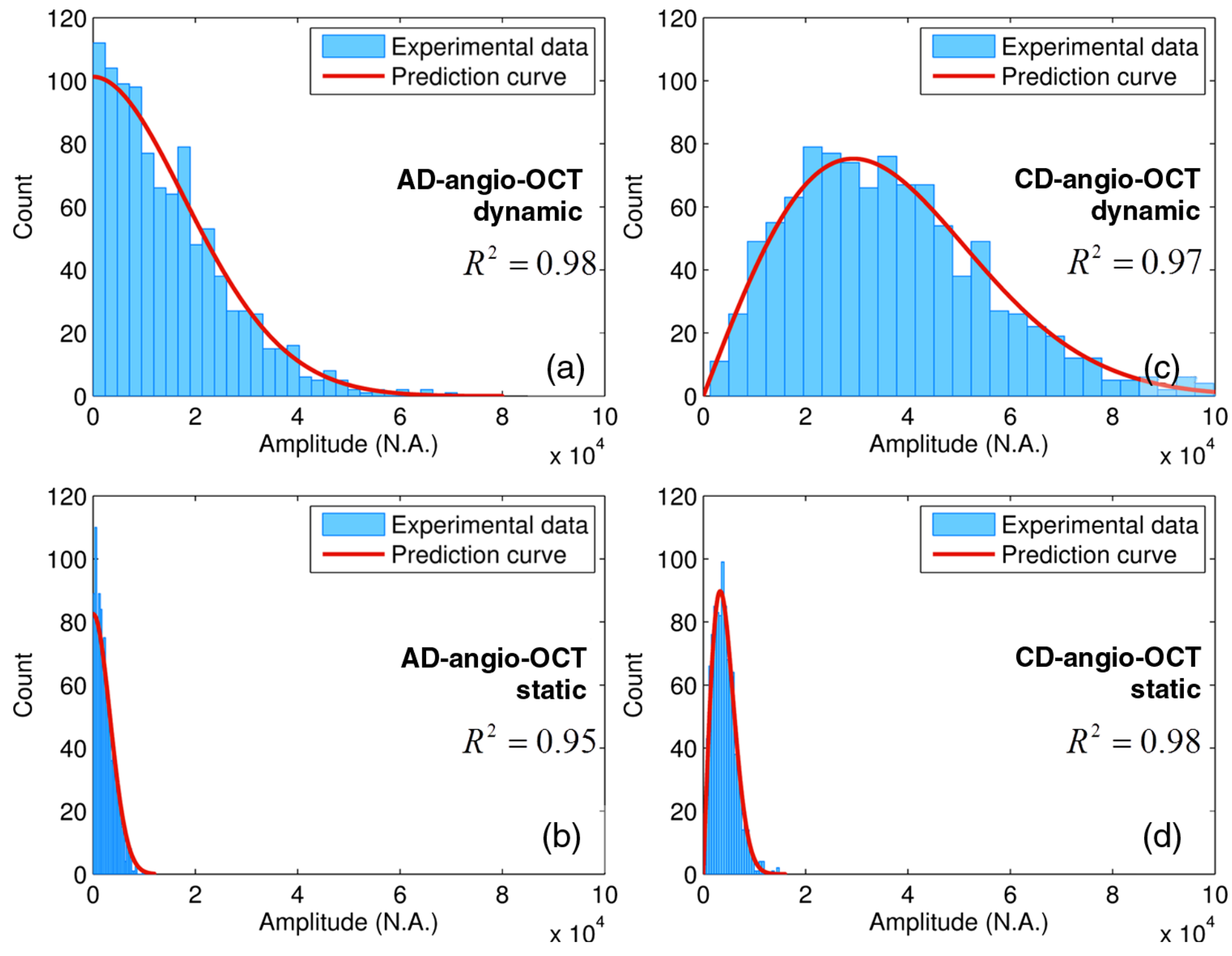

**Fig. 8** Statistics of the Angio-OCT signals in mouse brain *in vivo* imaging. (a, b) The statistics of the dynamic and static signals in AD-Angio-OCT, respectively. (c, d) The statistics of the dynamic and static signals in CD-Angio-OCT, respectively. The R-square statistic ($R^2$) is higher than 0.95.

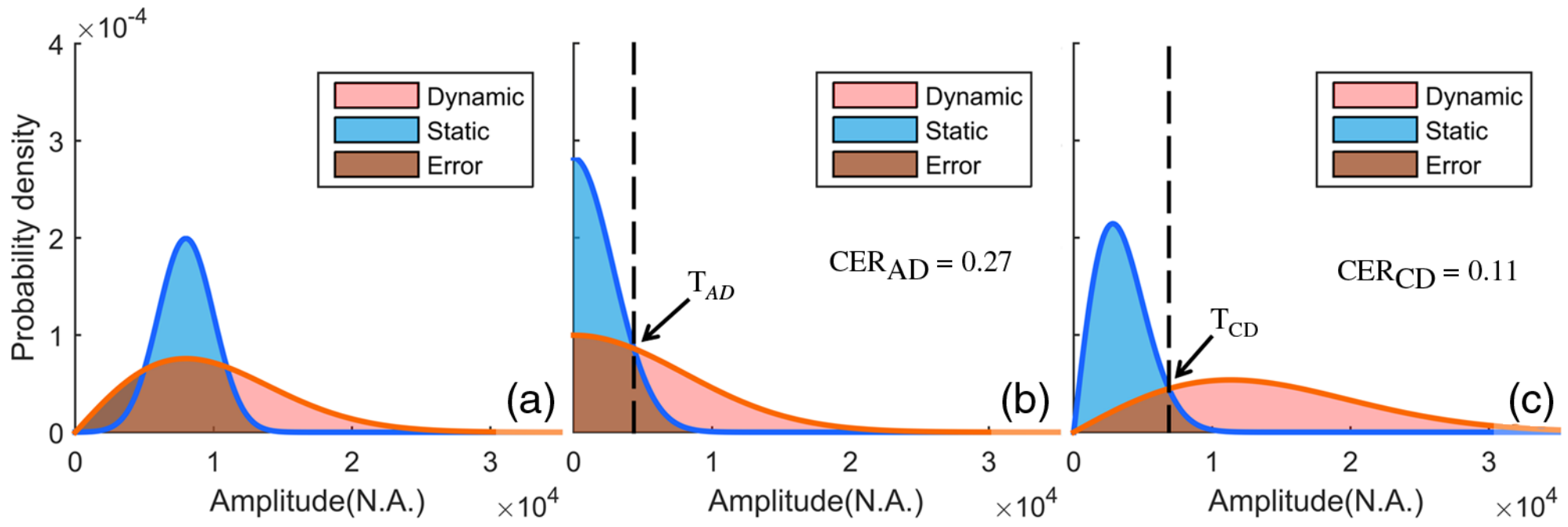

**Fig. 9** Normalized statistical distributions in OCT: (a) AD-Angio-OCT, (b) CD-Angio-OCT, and (c) $\sigma_d/\sigma_s = 4$.

The proposed statistical model can be used for guiding the threshold determination in Angio-OCT. Currently, the threshold is set empirically, and the signals above the threshold are classified as the dynamic regions. The ratio of the misclassified signals can be defined as the classification error rate (CER), i.e., dynamic signals below the threshold plus the static signals above it. In Angio-OCT, the minimal CER is determined by the residual overlap between the distributions of dynamic and static signals. Accordingly, the cross point of the two distribution curves can be considered as the optimal threshold. Any offset from the optimal value would lead to an increased CER. The optimal thresholds in the AD-Angio-OCT ($T_{\mathrm{AD}}$) and the CD-Angio-OCT ($T_{\mathrm{CD}}$) should meet the following conditions:

$$\begin{cases} f_{|A_{\mathrm{ADs}}|}(T_{\mathrm{AD}}) = f_{|A_{\mathrm{AD}d}|}(T_{\mathrm{AD}}) & \text{in AD} \\ f_{A_{\mathrm{CDs}}}(T_{\mathrm{CD}}) = f_{A_{\mathrm{CD}d}}(T_{\mathrm{CD}}) & \text{in CD.} \end{cases} \tag{26}$$

Substituting Eqs. (16), (20), (24), and (25) into the expressions above, we obtain:

$$T_{\mathrm{AD}} = 2\sigma_d\sigma_s\sqrt{\frac{1}{\sigma_d^2 - 2\sigma_s^2}\,\ln\left(\frac{\sigma_d}{\sqrt{2}\sigma_s}\right)}, \tag{27}$$

$$T_{\mathrm{CD}} = 2\sigma_d\sigma_s\sqrt{\frac{2}{\sigma_d^2 - \sigma_s^2}\,\ln\left(\frac{\sigma_d}{\sigma_s}\right)}. \tag{28}$$

Accordingly, the thresholds of the AD- and CD-Angio-OCT are determined as indicated by the dashed lines in Fig. 9. Then the minimal CER of the AD-Angio-OCT ($\mathrm{CER_{AD}}$) and the CD-Angio-OCT ($\mathrm{CER_{CD}}$) can be calculated:

$$\begin{aligned} 2\cdot\mathrm{CER_{AD}} &= \int_{T_{\mathrm{AD}}}^{\infty} f_{|A_{\mathrm{ADs}}|}(a_{\mathrm{ADs}})\mathrm{d}a_{\mathrm{ADs}} \\ &\quad + \int_0^{T_{\mathrm{AD}}} f_{|A_{\mathrm{AD}d}|}(a_{\mathrm{AD}d})\mathrm{d}a_{\mathrm{AD}d} \\ &= 1 + \mathrm{erf}\left[\sqrt{\frac{2}{k^2-2}\ln\left(\frac{k}{\sqrt{2}}\right)}\right] - \mathrm{erf}\left[\sqrt{\frac{k^2\ln(k)}{k^2-1}}\right], \end{aligned} \tag{29}$$

$$\begin{aligned} 2\cdot\mathrm{CER_{CD}} &= \int_{T_{\mathrm{CD}}}^{\infty} f_{A_{\mathrm{CDs}}}(a_{\mathrm{CDs}})\mathrm{d}a_{\mathrm{CDs}} + \int_0^{T_{\mathrm{CD}}} f_{A_{\mathrm{CD}d}}(a_{\mathrm{CD}d})\mathrm{d}a_{\mathrm{CD}d} \\ &= 1 - \exp\left[-\frac{2\ln(k)}{k^2-1}\right] + \exp\left[-\frac{2k^2\ln(k)}{k^2-1}\right], \end{aligned} \tag{30}$$

where we define $k = \sigma_d/\sigma_s$. In practice, a set of training data can be collected using MB-mode scanning protocol in the region of interest. The empirical threshold is first used to separate the dynamic and static regions. Based on the dynamic and static data, the parameters of $\sigma_d^2$ and $\sigma_s^2$ are learned, and the theoretical thresholds are determined. Initial proof of concept of the threshold determination was validated in this study. As shown in Figs. 3 and 6, the theoretical thresholds work well in the homogeneous tissues. Tissues of different scattering properties correspond to different parameters $\sigma_d^2$ and $\sigma_s^2$. In Fig. 6, the cortex was used for parameter learning and threshold determination, and consequently there exist apparent classification errors in the cranium using the threshold of cortex.

Although it has been recognized that the method using the complex-valued signals offers higher motion-contrast by combining both the amplitude and phase information, the performance of the CD- and AD-Angio-OCT can be further understood from the theoretical model. As reported in Fig. 10, the CD-Angio-OCT shows a lower CER in most situations, i.e., a superior motion-contrast, which can be confirmed in Figs. 3(b), 3(c), 6(b), and 6(c) as indicated by the yellow arrows. In Fig. 6, the averaged CERs of the AD and CD methods are 0.26 and 0.12 in the region of cortex, respectively. However, it should be noted that the CD algorithm is extremely sensitive to the phase fluctuation, and consequently poses a high requirement for the system phase stability and a large computational load for the phase compensation. The phase compensation works well on the situations, such as flow phantoms and stable animal models, but it is challenging in the clinical circumstances. Fortunately, several motion–tracking techniques have been developed in ophthalmic OCT systems for motion correction.

Averaging is widely used in Angio-OCT for high contrast. The developed model is helpful for guiding the design of the averaging approaches. According to the model, averaging of independent angiograms offers reduced CER and improved motion-contrast. Assuming two independent angiograms with the PDF $f_{\mathrm{AngioOCT}}$ expressed by Eqs. (16), (20), (24), and (25), the PDF of the averaged Angio-OCT signals $f_{\overline{\mathrm{AngioOCT}}}$ has a simple relation with the PDF of the original signals:

$$f_{\overline{\mathrm{AngioOCT}}} = f_{\mathrm{AngioOCT}} * f_{\mathrm{AngioOCT}}, \tag{31}$$

where $*$ represents the convolution computation. Figure 11 reports the normalized statistical distributions of the averaged

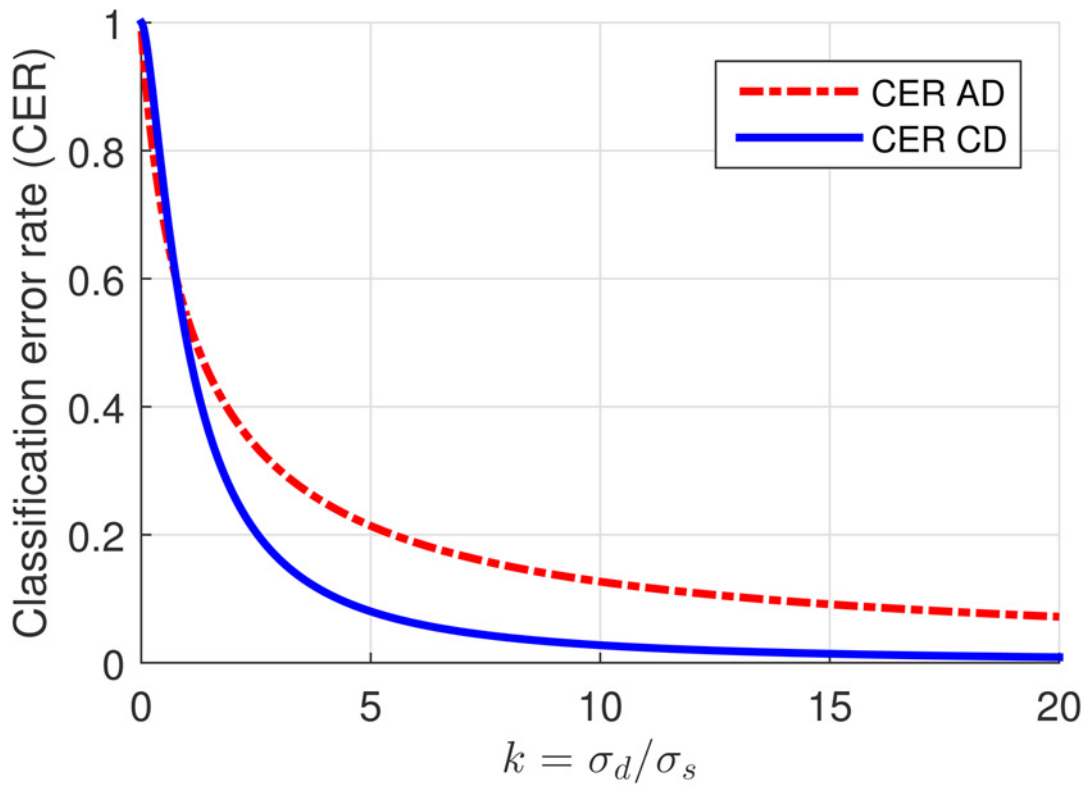

**Fig. 10** Numerical comparison of classification error rate (CER) between AD-Angio-OCT and CD-Angio-OCT. Complex differential (CD) method has a lower CER than AD.

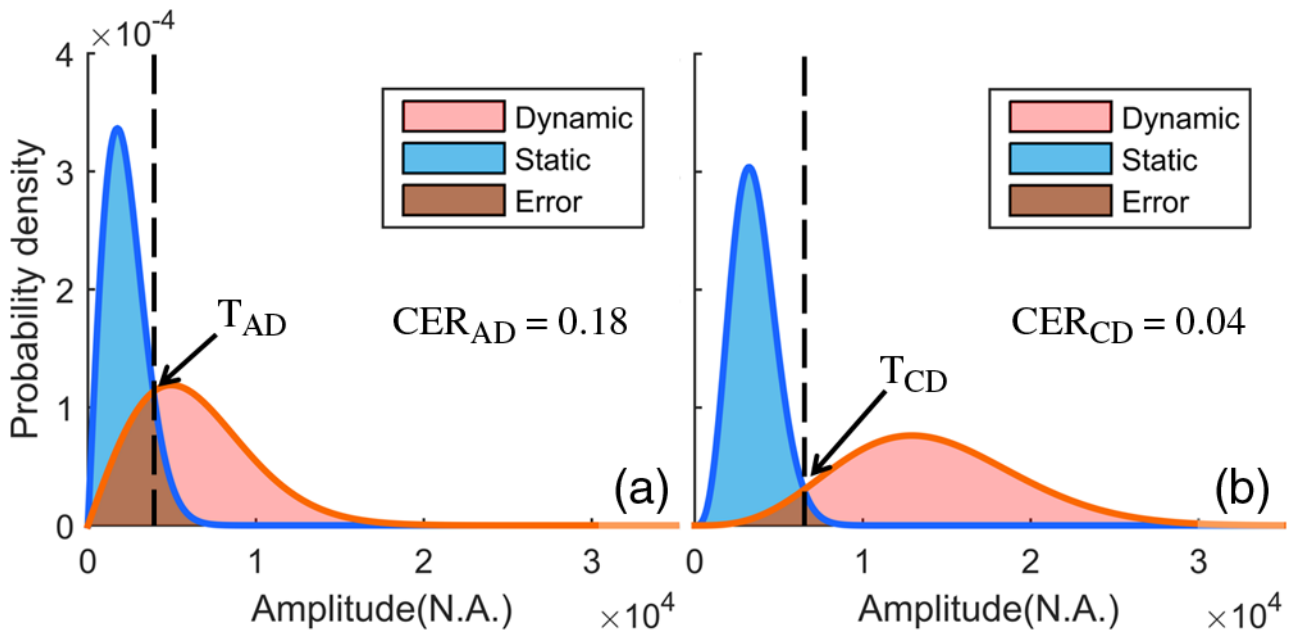

**Fig. 11** Normalized statistical distributions of the averaged (a) AD-Angio-OCT and (b) CD-Angio-OCT signals. $\sigma_d/\sigma_s = 4$.

Angio-OCT signals. Compared with the original distribution in Figs. 9(b) and 9(c), the averaged signals show a lower CER. If the angiograms are totally dependent, no contrast improvement can be obtained. Because the Angio-OCT signals of either the dynamic or static regions are random in the time dimension, repeated angiograms with a large time interval ($T$) can be considered as totally independent and can be used for averaging. However, the repeated imagings are performed in B-scans in typical Angio-OCT,[10,13,16] and consequently, the large time interval leads to an increased imaging time which is not desired due to the influence of bulk motion. In contrast, similar to the averaging approaches used for speckle reduction,[31] independent angiograms can also be achieved by methods such as wavelength diversity, angular diversity, and polarization diversity, and it can be explained in theory that the split-spectrum algorithm offers improved contrast.[10]

In Eqs. (13), (22), and (23), it is assumed that the dynamic signals in the $n$'th and $(n+1)$th B-frames are totally uncorrelated and independent under the condition of a sufficient time interval ($t$). Typically, the time interval ($t$) is determined by the B-frame rate in the interframe Angio-OCT,[10,13,16] which is 5.3 ms in our system. Such a time interval is sufficient for the fast blood flow, but not for the slow one. Taking the approximation where the lateral resolution of OCT is 15 $\mu$m, and the velocity of RBC in the capillaries is 1 mm/s,[32] the required $t$ is around 15 ms. According to the proposed model, in spite of being sensitive to the slow motion, the Angio-OCT with a short time interval would result in an increased CER and limited motion-contrast. Thus, there exists a tradeoff between the imaging speed and motion-contrast in Angio-OCT.

Angio-OCT suffers from shadow artifacts extending below the vessels, and the artifacts frustrate the automated 3-D analysis of vascular networks.[1] Due to the forward scattering of RBC, the static signals below the vessels are influenced by the dynamic multiple-scattered signals from the blood flow, and present a Rician distribution.[30] The statistical differences between the shadow and flow areas can be analyzed, which may be helpful for suppressing the shadow artifacts.

Although the mathematical derivation is focused on the AD and CD Angio-OCT in this study, it can be transferred to the phase-based methods. According to Eqs. (5) and (9), the temporal PDFs of phase $\theta_d$ and $\theta_s$ follows uniform and Gaussian distributions, respectively.[24] Then the statistics of the phase-based Angio-OCT can be deduced.

There are several limitations in the current model. First, in the dynamic regions, it is assumed that a large number of scattering RBCs are randomly distributed within the OCT voxel of interest, and the interframe OCT signals contributed by RBC are totally independent. The assumption can be well satisfied in the large blood vessels, but not in the capillaries. In the capillaries, RBC flow one by one at a slow speed (<1 mm/s).[32] The temporal statistics of the capillary signals in Angio-OCT will be investigated in future study. Second, in the static regions, we assume that the OCT scattering signal is a strong phasor and remains constant in all the B-frame for a given spatial point, referring to Eqs. (17), (22), and (23). When the bulk motion happens, there exists a relative change of the strong phasor. In particular, the boundaries of the layered tissues have great changes in both amplitude and phase. As shown in Figs. 6(b) and 6(c), obvious artifacts can be observed in the boundaries of the cranium, and the CD method shows more artifacts in the boundaries due to the considerable phase changes.

## 6 Conclusions

Based on the model of random phasor sums, the temporal statistics of the complex-valued OCT signals were mathematically described. Using mathematical transformations and reasonable approximations, the temporal statistics of AD- and CD-Angio-OCT signals were derived and were found to obey different statistical distributions. The theories were further validated through both the flow phantom and live animal experiments. Using the model developed in this work, the origin of the motion–contrast in Angio-OCT is mathematically explained, and the possible implications in the improvement of motion–contrast are further discussed, including the threshold determination and its residual classification error, averaging method, and scanning protocol. The CD-Angio-OCT shows a lower CER than the AD method when the phase compensation works well. The proposed mathematical model of Angio-OCT signals can aid in the optimal design of the system and associated algorithms.

## *Acknowledgments*

We acknowledge financial supports from National Natural Science Foundation of China (Nos. 61475143, 11404285, 61335003, 61327007, and 61275196), Zhejiang Provincial Natural Science Foundation of China (No. LY14F050007), National Hi-Tech Research and Development Program of China (No. 2015AA020515), Zhejiang Province Science and Technology Grant (No. 2015C33108), Fundamental Research Funds for the Central Universities (No. 2014QNA5017), and Scientific Research Foundation for Returned Scholars, Ministry of Education of China.

Biographies for the authors are not available.